\pdfoutput=1
\def\be{\begin{equation}}
\def\ee{\end{equation}}

\documentclass[showpacs,preprintnumbers,amsmath,amssymb,prl]{revtex4}

\usepackage{graphicx}
\usepackage{dcolumn}
\usepackage{bm}
\usepackage{subfigure}

\begin{document}

\title{Nonequilibrium Forces Between Neutral Atoms Mediated by a Quantum Field}

\author{Ryan O. Behunin$^1$ and Bei-Lok Hu$^{1,2}$}

\affiliation{$^1$ Maryland Center for Fundamental Physics, $^2$ Joint
Quantum Institute, \\ University of Maryland, College Park, Maryland,
20742}

\date{Feb 12,2010}

\begin{abstract}                                                                                                                                 

We study all known and as yet unknown forces between two neutral
atoms, modeled as three dimensional harmonic oscillators, arising
from mutual influences mediated by an electromagnetic field but not
from their direct interactions. We allow as dynamical variables the
center of mass motion of the atom, its internal degrees of freedom
and the quantum field treated relativistically. We adopt the method
of nonequilibrium quantum field theory which can provide a first
principle, systematic and unified description including the intrinsic
field fluctuations and induced dipole fluctuations. The inclusion of
self-consistent back-actions makes possible a fully dynamical
description of these forces valid for general atom motion.  In thermal
equilibrium we recover the  known forces -- London, van der Waals and
Casimir-Polder forces -- between neutral atoms in the long-time limit
but also discover the existence of two new types of interatomic
forces. The first, a `nonequilibrium force', arises when the field
and atoms are not in thermal equilibrium, and the second, which we
call an `entanglement force', originates from the correlations of the
internal degrees of freedom of entangled atoms.

\end{abstract}
\pacs{03.65.Yz,  11.10.Wx, 34.35.+a, 31.15.xk}

\maketitle

\section{Introduction}

In these notes we delineate a new approach to the study of forces
between two (but not limited to two)  neutral atoms mediated by a
quantum (in this case, electromagnetic) field based on nonequilibrium
quantum field theory \cite{NEqQFT}. This theory is ostensibly very
different from the usual approaches researchers in atomic, molecular
and optical (AMO) physics are familiar with in the treatment of
atomic-optical systems, and it may at first sight appear to be too
cumbersome or complicated to be necessary. However, with the advances
of sophisticated and highly controllable experiments in AMO physics
made possible by new high-precision instrumentation applied to cold
atoms in optical lattices (see, e.g., the experiments and the
theoretical analysis of  \cite{Rey04}) or cavities (with the
capability of tracking atoms in real time \cite{Orozco}) or
nanoelectromechanical systems  we are entering an age where
traditional theories will soon become inadequate. The present method
we introduce has the advantage that it is the amalgamation of both
quantum field theory and nonequilibrium statistical mechanics, the
former is required for quantum field (customarily referred to as
retardation, but there are more involved) effects, the latter for treating processes involving   
quantum dissipation and noises. Not only can this method reproduce
all the effects and forces known in the last century as detailed
below \cite{Lon,DLP, CasPol} , it can also deal with phenomena and
processes more recently brought to central attention from quantum
foundational and information processing issues, such as quantum
decoherence and entanglement dynamics, including non-Markovian
processes (those carrying memories) which invariably will appear when
back-action is taken into consideration. Since this method can treat
quantum back-action and feedback in a self-consistent manner, it is
uniquely adept to quantum control considerations \cite{QControl}.

Unlike most known treatments of systems in near-equilibrium
conditions based on linear or nonlinear response theory this is a
fully nonequilibrium dynamical description of the atoms' motion
\footnote{One simple way to tell the difference is whether
temperature is used ab initio or whether the system remains
stationary.}. Although our attention is focussed on the forces
between two neutral atoms derived by finding their relative
trajectories (which are determined self-consistently by the atoms'
interaction with the field and with each other via the field and the
field's fluctuations in the presence of the atoms), one can also
treat radiation reaction, dissipation, and fluctuation phenomena with
non-Markovian behaviors \cite{BH09}. As this paper will hopefully
illustrate a small initial investment into this new method can pay
bountifully.

We consider an assembly of $n$ neutral atoms (labeled by $a = 1,
...n$) and model the internal degrees of freedom (idf) of the $a$th
atom by a three dimensional harmonic oscillator with coordinates
$\vec{Q}_a$, (thus describing the atom's spontaneous and stimulated
emissions and absorption while interacting with a field).  The atoms
interact with an electromagnetic field (from near-field Coulomb force
to far-field radiation) with vector potential $A^\mu$ through a
dipole interaction, but not directly with one another. The force
between them arises through field-mediated mutual influences. The
non-relativistic trajectory of the $a$th atom is described by
$\vec{z}_a$ which, unlike in most previous treatments, is a dynamic
variable (not prescribed) determined self-consistently by a
negotiation amongst all the other variables $(\vec{Q}_a, A^\mu)$. Our
interest in this paper is primarily focussed on the center of mass
motion of each atom and not on the microscopic details of the other
variables. The open quantum systems \cite{qos} approach can
efficiently isolate the desired  information about the atom's
trajectory through a succession of coarse-graining procedures, as
detailed below, which take into account the overall effects
(back-action) of the remaining variables. Using the influence
functional method we can incorporate the effects of the microscopic
physics of the field and the atom's idf and derive an effective
equation of motion for the atom's trajectory from which the force
between the two atoms can be extracted by appealing to Newton's
Second Law.

In this setup the dipole moment of the atom modeled by an oscillator
is not permanent but only instantaneously non-vanishing.
The uneven distribution of charge in the atom comes from two effects.
First, semi-classically speaking, the magnitude and direction
of the nuclear-electron separation (which is proportional to the
 dipole moment of the atom) will unpredictably vary in time
even in the absence of quantum fields,
and second, in the presence of
quantum fields the atom is polarized by electric field fluctuations.
 Dipole moment fluctuations are the source of the interaction among neutral atoms
and are responsible for two types of forces arising from distinctly
different physical origin.

\subsubsection{Intrinsic Fluctuation Force}                                                                   

In the quantum-field conception of a neutral atom the electronic
wavefunction surrounding the nucleus has a fluctuating component,
modeled in our approach by a quantum mechanical harmonic oscillator.
%
%
As a whole the atom will always remain neutral. However,  in time
$\textit{intrinsic fluctuations}$ of the oscillator, due to its
quantum nature,  lead to an uneven local distribution of charge in an
otherwise (globally) neutral atom which gives rise to an
instantaneous dipole moment that couples to the attending electromagnetic field.
%
Radiation traveling away from the first (fluctuating)
atom carries information about the orientation of its dipole (at the
time of emission in the past) which eventually reaches and polarizes
the second atom.
The second atom's response to the field leads it
also to produce a time-varying electric field that travels
back to the first atom and is correlated with the activities of
the fluctuating atom's idf,  leading to nonvanishing interaction
energy.
One can think of the second atom as a transponder which
receives a signal from the fluctuating atom and then rebroadcasts it.
In this analogy the fluctuating atom will receive a signal
reflected from the transponder atom which encodes its own history.

This is easily conceptualized if we consider the atoms to be so close
that the light transit time between them is much greater than all other characteristic
time scales governing the dynamics. In such a case
the retarded electric field is well approximated by the
 the electrostatic field.
Thus, an intrinsic
fluctuation of the idf of one atom will source a static dipole electric field seen by the second atom.
The second atom is polarized by this external field
leading it too to source a dipole field felt by the fluctuating atom.
This process leads to an energetically
favorable arrangement of the two atom's dipole moments
which gives rise to the attractive force between them.

This type of force due to intrinsic fluctuations in the neutral
atoms' dipole moments contain two well-known forces: 1) the van der
Waals force, usually used to describe all interactions between
neutral atoms and molecules categorically, 
and
2)  the London force which arises from the Coulombic interaction
between atoms without permanent multipole moments and without the
consideration of retardation effects (as Casimir-Polder force does).
We refer to forces of this type as \textbf{intrinsic fluctuation
forces}.

\subsubsection{Induced Dipole Force}

It goes without saying that the quantum field itself possesses
intrinsic fluctuations. Any instantaneously generated local electric
field 
will  $\it{induce}$ non-vanishing dipole moments
in both atoms. We classify the interaction of
dipole moments induced by the fluctuations of the quantum
field as \textbf{induced dipole forces}. We suggest making a clean
separation between forces arising from intrinsic (before) and induced
(here) fluctuations of the dipole moment of a neutral atom because
the physical processes produce quite distinct results, as shown in
later sections.

The physical origin of this component of the force is the spatial
correlation of field fluctuations. Any given field fluctuation
will induce correlated dipole moments for the two atoms, much like a
long wavelength water wave on the ocean will raise and lower two
nearby buoys in phase. The excitation of the dipoles by the field
will lead to radiation that contains information about the emitter.
When the radiation from one atom reaches the other the correlation
between the induced motion of each dipole moment at the time of
emission, and subsequent communication of that motion via radiation
 leads to a nonvanishing interaction energy.

A well known force of this nature is that of Casimir-Polder (CP)
\cite{CasPol} who included considerations of the quantum nature of the
field. This CP force (there is also the CP force between an atom and
a mirror which will be treated in our second paper) is a
generalization of the London description including retardation
corrections as well as effects of field quantization -- quantization
being what imbues the field with its own intrinsic fluctuations.

\subsubsection{Coarse-graining and Back-action}

For a description of the forces between two atoms we need to know
only the averaged effect of the quantum field and the oscillator's
idf on the atom's trajectory, their details are not of great concern
in this quest.   Imagine the transition amplitude for the
$\it{total}$ system to evolve from some initial state, $\left|
\vec{z}_{in}, \varphi_{in} \right>$ to some final state $\left|
\vec{z}_{out}, \varphi_{out} \right>$ in time $T$ where $\vec{z}$
labels the atom's position and $\varphi$ is a collective label of the
state of all the remaining (environment) variables in the total
system. Our primary interest is the time development of the atom's
center of mass for which the field and its interaction with the
atom's idf plays a central role through processes like dissipation
and radiation reaction. For a given final position of the center of
mass there can be many consistent final field and oscillator states,
likely unobservable.
Summing over all final environment states compatible with the atom's
motion 
is $\it{necessary}$ when we are ignorant of the final state of the
environment whether we choose to ignore those details or they are not
measurable. This leads to an effective transition amplitude for the
trajectory of the atom $\it{alone}$ where all environmental effects
on the trajectory have been taken into account.   Carrying out this
process of $\it{coarse-graining}$ where the final field and
oscillator states are traced over leads to an effective action that
self-consistently accounts for all back-action of the field and the
atom's idf on the atom's trajectory. The equation of motion for the
atom, and thus the atom-atom force can be obtained through a
variation of this action \cite{BH09}.

As we shall show the present formulation goes beyond previous work in
that we can derive the forces between two atoms for fully dynamical
and under nonequilibrium conditions. When the spacing between the
atoms is held fixed we recover the well-known London and CP forces.
For the case where the atoms and field are not in thermal equilibrium
we find a novel far field scaling for the induced dipole force
diminishing as $1/z^3$ rather than $1/z^8$, and for the case when the
two atom's are entangled we find a novel near field scaling that
enters at second order in perturbation theory as $q^2/z^2$ as opposed
to the standard $q^4/z^7$ where $z$ quantifies the interatomic
distance and $q$ the electronic charge. To the best of our knowledge
this nonequilibrium force and the entanglement force behavior have
not been reported in the literature.

This paper is organized as follows:  In Sec. 2 we introduce the
microscopic details of the system by defining the action describing
the dynamics of the entire system. The worldline influence functional
method is adopted where the environment degrees of freedom (field +
oscillator) are traced over to find the time evolution of the reduced
density matrix. The equations of motion for the atomic trajectories
are then obtained from the saddle points of the reduced density
matrix. In Sec. 3 the explicit form for the atom-atom force is
computed, and physical origin of each component is explained. In Sec.
4 an initially entangled state for the two oscillators is considered.
It is found that the correlation among the oscillator coordinates
leads to a contribution to the atom-atom force that enters at second
order in perturbation theory. Sec. 5 concerns the possibility of
detecting the nonequilibrium atom-atom force (when the field and
atoms are not in thermal equilibrium) which is new and the novel
entanglement force we discovered as outlined in Sec. 4.

\section{The Model}

 We describe the microphysical degrees of freedom of the entire system through the following action

\be
S[\vec{Q}_a,\vec{z}_a,A^\mu]= \sum_{a}
(S_Q[\vec{Q}_a]+S_Z[\vec{z}_a]+S_{int}[\vec{Q}_a,\vec{z}_a,A^\mu]
)+S_{E}[A^\mu]
\ee
where the sum is over all atoms. The action
describing the dynamics of the oscillator is given by

\be S_Q[\vec{Q}_a]=\frac{\mu_a}{2}\int
d\lambda[\dot{\vec{Q}}_a(\lambda)^2-\Omega^2_a\vec{Q}_a(\lambda)^2]
\ee where $\mu_a$  is the $a$th oscillator's reduced mass and
$\lambda$ its worldline parameter, $\Omega_a$ being its natural
frequency. The electomagnetic field action is given by

\be S_{E}[A^\mu]=-\frac{1}{4}\int d^4x F_{\mu\nu} F^{\mu\nu} \ee (the
subscript $E$ stands for the electric field) where $A^\mu$ is the $4$-vector
potential and $F_{\mu \nu}=\partial_{\mu} A_{\nu}-\partial_{\nu}
A_{\mu}$ is the field strength tensor. The action for the center of
mass motion is \be S_Z[\vec{z}_a]=\int d\lambda \bigg[ \frac{M_a}{2}
{\dot{\vec{z}}_a}^2(\lambda)-V[\vec{z}_a] \bigg] \ee where $M_a$ is
the atom's total mass and $V[\vec{z}_a]$ is an external potential.

In the dipole approximation,  the potential energy for an atom
interacting with the photon field takes the form $-q \ \vec{Q}\cdot
\vec{E}[\vec{z}]$, where $q\vec{Q}$ is the atom's instantaneous
dipole moment and $\vec{E}$ is the electric field leading to the
interaction action $ S_{int}[\vec{Q}_a, \vec{z}_a , A_\mu]= q_a \int
d\lambda {Q}^i_a(\lambda) E_{i}[z^{\mu}_a(\lambda)] $. Above, $q_a$,
quantifies the coupling of the $a$th atom to the field. [Greek
indices will refer to spacetime components of a four-vector, zero
referring to time, and Roman indices refer to spatial components
where we will exclusively use the letters  $\{i,j,k\}$ to avoid
confusion with the letter $a$ used to label atoms. Contraction of
four-vectors is undertaken with the Minkowski metric with (-,+,+,+)
signature, and the Einstein summation convention is used throughout.


\subsubsection{Worldline Influence Functional }

 Assume that at time $t_{in}$ the quantum statistical state of the oscillators, trajectory and field is described by a density operator $\hat{\rho}(t_{in})$. This state is unitarily evolved from the initial time $t_{in}$ to a later time $t>t_{in}$, and can be expressed in terms of path integrals by considering matrix elements in an appropriate basis.

To isolate the influence of the field on the dynamics of the
atom we coarse-grain over the field variables to construct the
field-reduced density matrix,
$
\rho_r(\vec{Q}_a,\vec{Q}'_a;\vec{z}_a,\vec{z}'_a;t)=\int dA^{\mu} \ {\rho}(\vec{Q}_a,\vec{Q}'_a  ;\vec{z}_a,\vec{z}'_a; A^\mu, A^\mu;t).
$
By assuming that the field is initially uncorrelated with the other degrees of freedom the reduced density matrix takes the form,

\begin{eqnarray}
\label{rhor}
\rho_r(\vec{Q}_a,\vec{Q}'_a;\vec{z}_a,\vec{z}'_a;t)= \prod_a \int d \vec{Q}_{in,a}   \ d\vec{Q}'_{in,a}
\int d\vec{z}_{in,a}  \ d\vec{z}'_{in,a}
\int_{\vec{Q}_{in,a}}^{\vec{Q}_a}\mathcal{D}\vec{Q}_a
\int_{\vec{Q}_{in,a}'}^{\vec{Q}'_a}\
 \mathcal{D}\vec{Q}'_a
\int_{\vec{z}_{in,a}}^{\vec{z}_a}\mathcal{D}\vec{z} _a
\int_{\vec{z}_{in,a}'}^{\vec{z}'_a}
\mathcal{D}\vec{z}' _a
\nonumber \\ \times e^{i(S_Q[\vec{Q}_a]+S_Z[\vec{z}_a]-S_Q[\vec{Q}'_a]-S_Z[\vec{z}'_a])}  \rho_{Qa}(\vec{Q}_{in,a},\vec{Q}_{in,a}';t_{in})
\rho_Z(\vec{z}_{in,a},\vec{z}_{in,a}';t_{in}) \mathcal{F}[{J}^{\mu-}, {J}^{\nu+}]
\end{eqnarray}
which introduces the influence functional (IF) $\mathcal{F}[{J}^{\mu-}, {J}^{\nu+}]$ \cite{FeyVer}. If the initial state of the field is Gaussian in field variables (which includes vacuum and thermal states) the influence functional can be calculated exactly for the dipole field interaction.
\begin{equation}
\label{IF}
\mathcal{F}[{J}^{\mu-}, {J}^{\nu+}]=\exp\bigg\{i\int d^4y\int d^4y' [J^{\mu-}(y)D^{ret}_{\mu \nu}(y,y')J^{\nu+}(y')+\frac{i}{4}J^{\mu-}(y)D^H_{\mu \nu}(y,y')J^{\nu-}(y')]\bigg\}
\end{equation}
Here the current density is

\begin{equation}
\label{CD}
J_\mu(x)=- \sum_a q_a \int d\lambda \kappa_{i\mu} \delta^4(x^\mu-z^\mu_a(\lambda))Q^i_a(\lambda),
\end{equation}
  $J^+=(J+J')/2$ and $J^-=J-J'$ are its semi-sum and difference, respectively, where prime distinguishes histories, and $\kappa_{i\mu}=\partial_i \eta_{0 \mu}-\partial_0 \eta_{i \mu}$ is a differential operator that relates the photon field to the electric field by contraction i.e. $E^i = \kappa^i_\mu A^\mu$.
$D^{ret}_{\mu \nu}(y,y')$ and $D^H_{\mu \nu}(y,y')$ are the retarded Green's function and Hadamard function for the field respectively. In the Feynman gauge they can be expressed in terms of the retarded, $D_{ret}$, and Hadamard, $D_H$, Green's function for a massles scalar field.
\begin{equation}
D^{ret}_{\mu \nu}(x,x')=\eta_{\mu \nu}D_{ret}(x,x')  \ \ \ \ \ \  D^{H}_{\mu \nu}(x,x')=\eta_{\mu \nu}D_{H}(x,x')
\end{equation}
At zero temperature they take on the explicit form
\begin{equation}
D_{ret}(x,x')=\frac{1}{4\pi}\theta(t-t')\delta(\sigma) \ \ \ \ \ \  D_{H}(x,x')=-\frac{1}{4\pi^2 \sigma}
\end{equation}
where $\sigma$ is Synge's worldfunction defined to be half the geodesic distance between the four-vectors $x$ and $x'$, $\sigma=(x-x')^2/2$.


\subsubsection{Oscillator-Reduced Influence Functional}

We isolate the net influence that the oscillator's idf $\vec{Q}_a$ and the
field $A^\mu$ have on the trajectory by tracing over all final
oscillator configurations $\rho_{or}(\vec{z}_a, \vec{z}'_a; t)= \int
d\vec{Q}_a \rho_r(\vec{Q}_a,\vec{Q}_a;\vec{z}_a,\vec{z}'_a;t)$
introducing the oscillator-reduced density matrix $\rho_{or}$.

\begin{eqnarray}
\label{ordm}
\rho_{or}(\vec{z}, \vec{z}'; t) =
\int d\vec{z}_{in,1}  \ d\vec{z}'_{in,1}
\int_{\vec{z}_{in,1}}^{\vec{z}}\mathcal{D}\vec{z} _1
\int_{\vec{z}_{in,1}'}^{\vec{z}'}
\mathcal{D}\vec{z}' _1
 e^{i(S_Z[\vec{z}_1]-S_Z[\vec{z}'_1])} \rho_Z(\vec{z}_{in,1},\vec{z}_{in,1}';t_{in}) \mathcal{F}_Z[\vec{z}_1,\vec{z}'_1]
\end{eqnarray}
All the effects of the environment are now packaged in the
oscillator-reduced IF, $\mathcal{F}_Z[\vec{z}^-,\vec{z}^+]$. The
development has been simplified by working in the rest frame of the
second atom in so doing  $\vec{z}_2$ is no longer treated as a
dynamical variable.

\begin{eqnarray}
\label{orIF1}
\mathcal{F}_Z[\vec{z}^-_1,\vec{z}^+_1]  =  \prod_a \int d\vec{Q}_a d\vec{Q}_{in,a}   d\vec{Q}'_{in,a}
\int_{\vec{Q}_{in,a}}^{\vec{Q}_a}  \mathcal{D}\vec{Q}_a
\int_{\vec{Q}'_{in,a}}^{\vec{Q}_a}
  \mathcal{D}\vec{Q}'_a
e^{i(S_Q[\vec{Q}_a]-S_Q[\vec{Q}'_a])}
 \rho_{Q_a}(\vec{Q}_{in,a},\vec{Q}'_{in,a} ;t_{in})  \mathcal{F}[J^{\mu-},J^{\nu+}]
\end{eqnarray}

To elucidate our approach we write (\ref{orIF1}) in a more
suggestive form

\begin{eqnarray}
\label{orIF}
\mathcal{F}_Z[\vec{z}^{+}_1,\vec{z}^{-}_1]= \mathcal{F}\left[\vec{z}^+_a, \vec{z}^-_a ; -i\frac{\delta}{\delta {\vec{j}_a}^+},-i\frac{\delta}{\delta {\vec{j}_a}^-}  \right]  \prod_{a}
F_{a}[\vec{j}_a^+,\vec{j}_a^-] \bigg|_{{j_a}^\pm=0}
\end{eqnarray}
which defines the IF for a three dimensional harmonic oscillator,
$F_{a}[\vec{j}_a^+,\vec{j}_a^-]$.
To bring $ \mathcal{F}[J^{\mu-},J^{\nu+}]$ out of
the path integrals in (\ref{orIF1}) $[Q^{k \pm}_a(\lambda)]^n$ is
replaced with functional derivatives on the IF for the harmonic
oscillators $\left(-i\frac{\delta}{\delta {j_a}^{\mp}_k(\lambda)}
\right)^n F_{a}[\vec{j}_a^+,\vec{j}_a^-] \bigg|_{{j_a}^{\pm}=0}$. The
explicit form for $F_{a}[\vec{j}_a^+,\vec{j}_a^-]$ is

\begin{equation}
\label{fo} F_{a}[\vec{j}_a^+,\vec{j}_a^-] =\int d{\vec{Q}_a } \int
d{\vec{Q}_{in,a}} d{\vec{Q}_{in,a}}  \
\rho_{Q_a}(\vec{Q}_{in,a},\vec{Q}_{in,a} ; t_{in} )
\int_{\vec{Q}_{in,a}}^{\vec{Q}_a}
\mathcal{D}\vec{Q}_a\int_{\vec{Q}_{in,a}}^{\vec{Q}_a}
\mathcal{D}\vec{Q}'_a  e^{iS_Q[\vec{Q}_a]-iS_Q[\vec{Q}'_a]+i\int
d\lambda [\vec{j}_a^+\cdot\vec{Q}_a^-+\vec{j}_a^-\cdot\vec{Q}_a^+]}
\end{equation}
In the above the dot product between current and oscillator
coordinate i.e. $\vec{j} \cdot \vec{Q}$ is taken with respect to a
three dimensional Euclidean metric. For a Gaussian initial state
(\ref{fo}) can be evaluated exactly
\begin{eqnarray}
\label{ } F_{a}[\vec{j}_a^+,\vec{j}_a^-] =\mathcal{N} \exp \bigg\{
i\int d\lambda d\lambda' [
\vec{j}_a^-(\lambda)\cdot\vec{j}_a^+(\lambda')g_{ret,a}(\lambda,\lambda')+\frac{i}{4}\vec{j}_a^-(\lambda)
\cdot\vec{j}_a^-(\lambda')g_{H,a}(\lambda,\lambda')] \bigg\}
\end{eqnarray}
where $g_{ret,a}(\lambda,\lambda')$ and $g_{H,a}(\lambda,\lambda')$ (expressed below at $T=0$) are the retarded and Hadamard Green's functions for a one dimensional harmonic oscillator with natural frequency $\Omega_a$, mass $\mu_a$, and $\mathcal{N}$ is a normalization constant.

\begin{equation}
\label{ }
g_{ret,a}(\lambda,\lambda')=\frac{1}{\mu_a \Omega_a}\theta(\lambda-\lambda')\sin \Omega_a(\lambda-\lambda')  \ \ \ \ \ g_{H,a}(\lambda,\lambda')=\frac{1}{\mu_a\Omega_a}\cos \Omega_a(\lambda-\lambda')
\end{equation}


\subsubsection{Decoherence and the Semi-Classical Limit}

The complex norm of (\ref{orIF1}) at leading order in a $\vec{z}_1^-$-expansion follows

\begin{eqnarray}
\label{}
| \mathcal{F}_Z[\vec{z}_1^-,\vec{z}_1^+] | = \exp\bigg\{-\int d\lambda d\lambda' \  z^{i-}_1(\lambda) N_{ij}(\lambda,\lambda') z^{j-}_1(\lambda')    \bigg\}.
\end{eqnarray}
where $N_{ij}$ is a symmetric positive definite kernel.
Thus, we observe that the off-diagonal elements of the density matrix in (\ref{rhor}) are strongly suppressed for large values of $\vec{z}^-=\vec{z}-\vec{z}'$ as is indicative of decoherence of the quantum trajectory \cite{HPZ1}.
Decoherence of the trajectory due to its interactions with the quantum
fluctuations of the environment and the internal degrees of freedom of the atoms permits the existence
of a semi-classical limit for the oscillator's path through space.
Using a saddle-point approximation to evaluate (\ref{ordm}) one can show
that the semi-classical dynamics is determined from the variation

\begin{equation}
\label{fd}
\frac{\delta S_{CGEA}[\vec{z}^{+}_1,\vec{z}^{-}_1]}{\delta z^{k-}_1(\tau)}\bigg|_{z^{k-}_1=0}=0 \Longrightarrow M \ddot{z}_k(\tau)=f_k(\tau)
\end{equation}
where the so-called coarse-grained effective action is given by
$S_{CGEA}[z_1^{i+},z_1^{i-}]=S_Z[\vec{z}_1]-S_Z[\vec{z}'_1]+S_{IF}[\vec{z}^+_1,\vec{z}^-_1]$,
and  $S_{IF}[\vec{z}^+_1,\vec{z}^-_1]=-i\ln
\mathcal{F}_Z[\vec{z}_1^-,\vec{z}_1^+]$ defines the  influence
action. The force acting on the trajectory due to its interactions
with the oscillators and field is given by

\begin{equation}
\label{inff}
f_k(\tau)=\frac{\delta S_{IF}[\vec{z}^{+}_1,\vec{z}^{-}_1]}{\delta z^{k-}_1(\tau)}\bigg|_{z^{k-}_1=0}.
\end{equation}
For general atom motion this force contains all known effects,
including the Lamb shift, radiation reaction, dissipation, and the
atom-atom force.

\section{Nonequilibrium Atom-Atom Force}

The suppression of the reduced density matrix for off-diagonal elements justifies an expansion of (\ref{orIF}) for small values of ${\vec{z}_1}^-$. 
The linear order term yields the influence force and is represented
by an infinite series in powers of the coupling. The local (spatially
independent) terms in this expansion lead to the aforementioned Lamb
shift, radiation reaction, and dissipation. The atom-atom force can
be obtained from this series by extracting the terms that depend upon
the spatial separation of the atoms.

To simplify the presentation we rewrite the influence functional for the atom's trajectory as $\mathcal{F}_Z= \left<    e^{i S_{eff}} \right>_o$
where the form of $S_{eff}=-i \ln \mathcal{F}$ can be taken from (\ref{IF})
\begin{equation}
\label{ }
S_{eff}= \int d^4 x \int d^4 x' [ J^{\mu-}(x) D^{ret}_{\mu \nu}(x,x') J^{\nu+}(x') +\frac{i}{4}  J^{\mu-}(x) D^{H}_{\mu \nu}(x,x') J^{\nu-}(x') ]
\end{equation}
and $\left< ...\right>_o$ is the expectation value with respect to
both oscillators under the condition of no interactions.

\begin{eqnarray}
\label{ }
\left< ...\right>_o= \prod_a \int d\vec{Q}_a d\vec{Q}_{in,a} d\vec{Q}'_{in,a} \ \rho_{Q_a}(Q_{in,a}, Q'_{in,a};t_{in}) \int_{\vec{Q}_{in,a}}^{\vec{Q}_a } \mathcal{D} \vec{Q}_a \int_{\vec{Q}'_{in,a}}^{\vec{Q}_a } \mathcal{D} \vec{Q}'_a e^{i(S_Q[\vec{Q}_a]-S_Q[\vec{Q}'_a])}   (...)
\end{eqnarray}

$S_{eff}$ is a quadratic function of the current density (\ref{CD})
which depends on a sum  of delta functions with support at each
atom's position. Thus, one can see that the cross terms $S_{cross}$
between the two atom's currents appearing in $S_{eff}$ will lead to
terms that depend upon their spatial separation.

\begin{equation}
\label{CR}
S_{cross}=\int d^4 x d^4 x' [ J_1^{\mu-}(x)
D^{ret}_{\mu\nu}(x,x') J_2^{\nu+}(x')+J_2^{\mu-}(x)
D^{ret}_{\mu\nu}(x,x') J_1^{\nu+}(x')+\frac{i}{2}J_1^{\mu-}(x)
D^{H}_{\mu\nu}(x,x') J_2^{\nu-}(x') ]
\end{equation}
$J_a^\mu$ refers to the  current density of the $a$th atom,
with $a=1$ referring to the distinguished atom, where all the
atom-atom forces we are studying here act upon.
$\left< S_{cross} \right>_o$, the expectation value of $S_{cross}$,
will vanish for initially uncorrelated and Gaussian oscillator states
because it is linear in the coordinate of each oscillator. Therefore
the leading order contribution to the atom-atom force will be
proportional to the square of $S_{cross}$ at order $q_1^2q_2^2$.
Expanding $\mathcal{F}_Z$ in powers of $S_{eff}$ we express the IF as
\begin{eqnarray}
\label{ }
\mathcal{F}_Z  =  e^{iS_{IF}[z^\pm]} \approx  1+(\text{self energy terms})-\frac{1}{2} \left< (S_{cross}[\vec{z}_a^\pm,\vec{Q}_a^\pm])^2\right>_o .
\end{eqnarray}
The leading order linear terms $\mathcal{O}(q^2_a)$ contain the
back-action of the field on the motion of the atom itself only. We
refer to them as``self energy terms", borrowing a terminology from
particle physics, and for a stationary atom these effects are
unimportant. We focus on the quadratic term which contains not only
higher order self energy type effects but also the leading order
contribution to the atom-atom force contained in $S_{cross}$.

Note that $S_{eff}$ contains a term linear in the retarded propagator
for the field.  It makes sense that the force will manifest as the
square of (\ref{CR}) because, as described in the introduction,
fluctuations in the dipole moment of one atom induce radiation that
travels to the other, influences its dynamics, and induces the other
atom to radiate. From a diagrammatic viewpoint this process requires
two propagators.

The leading order expression for the force can be derived from

\begin{equation}
\label{fAA}
f_k(\tau)=\frac{\delta S_{IF}[\vec{z}_1^\pm]}{\delta z^{k-}_1}\bigg|_{z^{k-}=0}\approx -\frac{i}{2} \frac{\delta }{\delta z_1^{k-}(\tau)}\left< (S_{cross}[\vec{z}^\pm_a,\vec{Q}^\pm_a])^2\right>_o \bigg|_{z^{k-}=0}
\end{equation}
where the first equality holds for a stationary trajectory (radiation reaction and dissipation vanish).

Carrying out the variation in (\ref{fAA}) yields three contributions
the first two being

\begin{equation}
\label{fA}
f^A_k(\tau)=\frac{1}{2}q_1^2 q_2^2
\int_{\lambda_i}^{\lambda_f} d\lambda \int_V d^4 x \int_V d^4 y \ g_{ret,1}(\tau, \lambda) E^{ij}_{ret}(z^\alpha(\lambda), y) G_H(x,y)  \partial_k(x') E^{ret}_{ij}(x', x)|_{x'=z^\alpha(\tau)}
\end{equation}

\begin{equation}
\label{fB}
f^B_k(\tau)=\frac{1}{2}q_1^2 q_2^2
 \int_{\lambda_i}^{\lambda_f} d\lambda \int_V d^4 x \int_V d^4 y \ g_{H,1}(\tau, \lambda) E^{ij}_{ret}(z^\alpha(\lambda), y) G_{ret}(x,y)  \partial_k(x') E^{ret}_{ij}(x', x)|_{x'=z^\alpha(\tau)}
\end{equation}
where $E_{ij}(x,x')$ is the dyadic electric field Green's function
$E_{ij}(x,x')=\text{Tr} \{ \hat{\rho}_E \hat{E}_i(x) \hat{E}_j(x')
\}$ which arises in our formalism through a contraction of the
operator $\kappa^i_\mu$ with the photon field. Note also that after
all functional derivatives are taken $z_1 \rightarrow z$,
$G_{ret}(x,x')$ and $G_{H}(x,x')$ are meant to generally represent
the retarded and Hadamard functions for whatever the atom interacts
with. For example to find the surface-atom
force the integration volume $V$ is taken to be the half space, and
the Green's functions describing the physics of the media occupying
that region are used. For the specific case of an atom located at
$\vec{z}_2$, $G(x,x')\propto g_2(t,t')
\delta^3(\vec{x}-\vec{z}_2)\delta^3(\vec{x}'-\vec{z}_2)$. More
general cases will be considered in a future paper.

\begin{figure}
\begin{center}
\mbox{\subfigure{\includegraphics[width=3in]{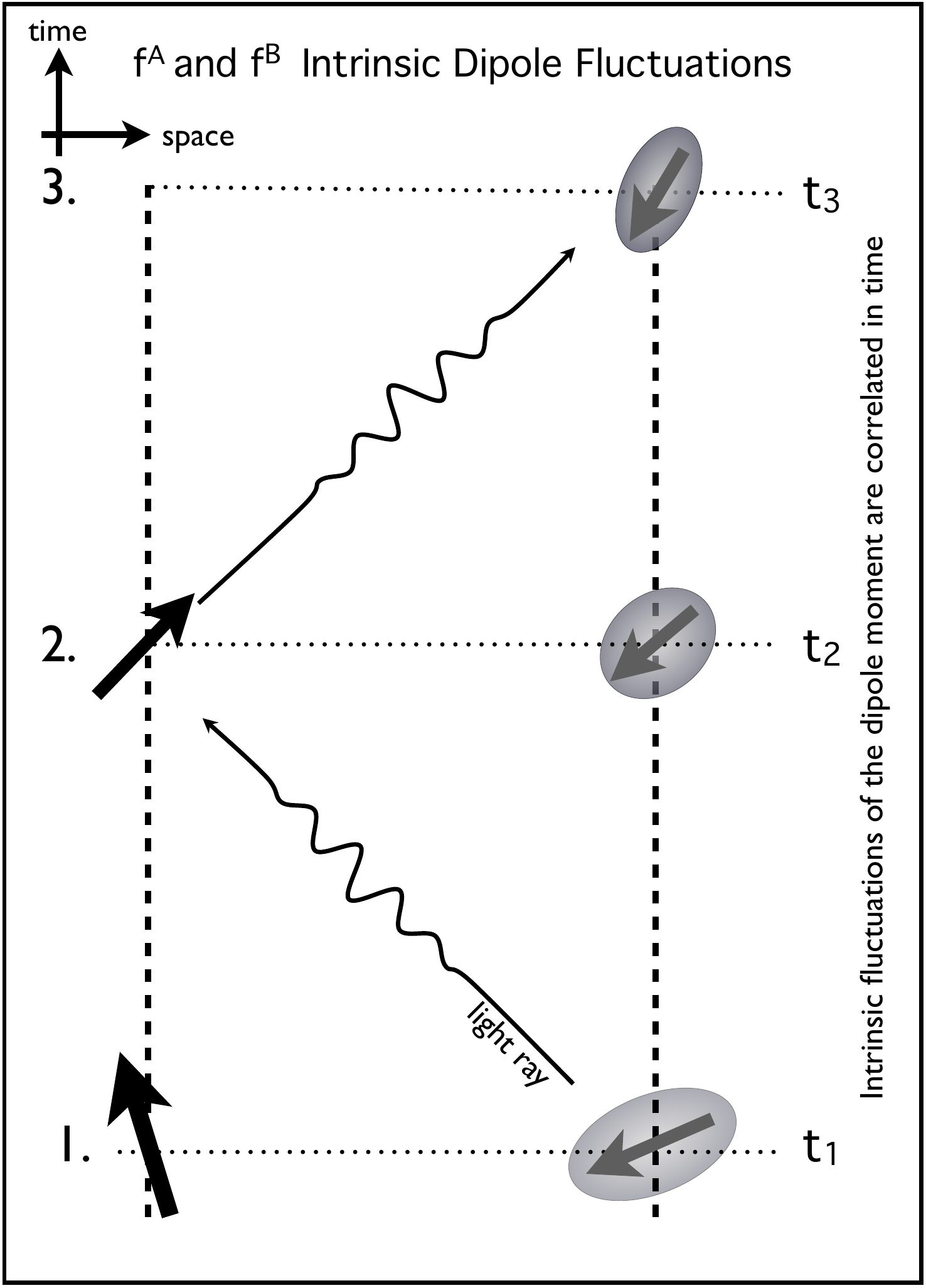}}\quad
\subfigure{\includegraphics[width=3in]{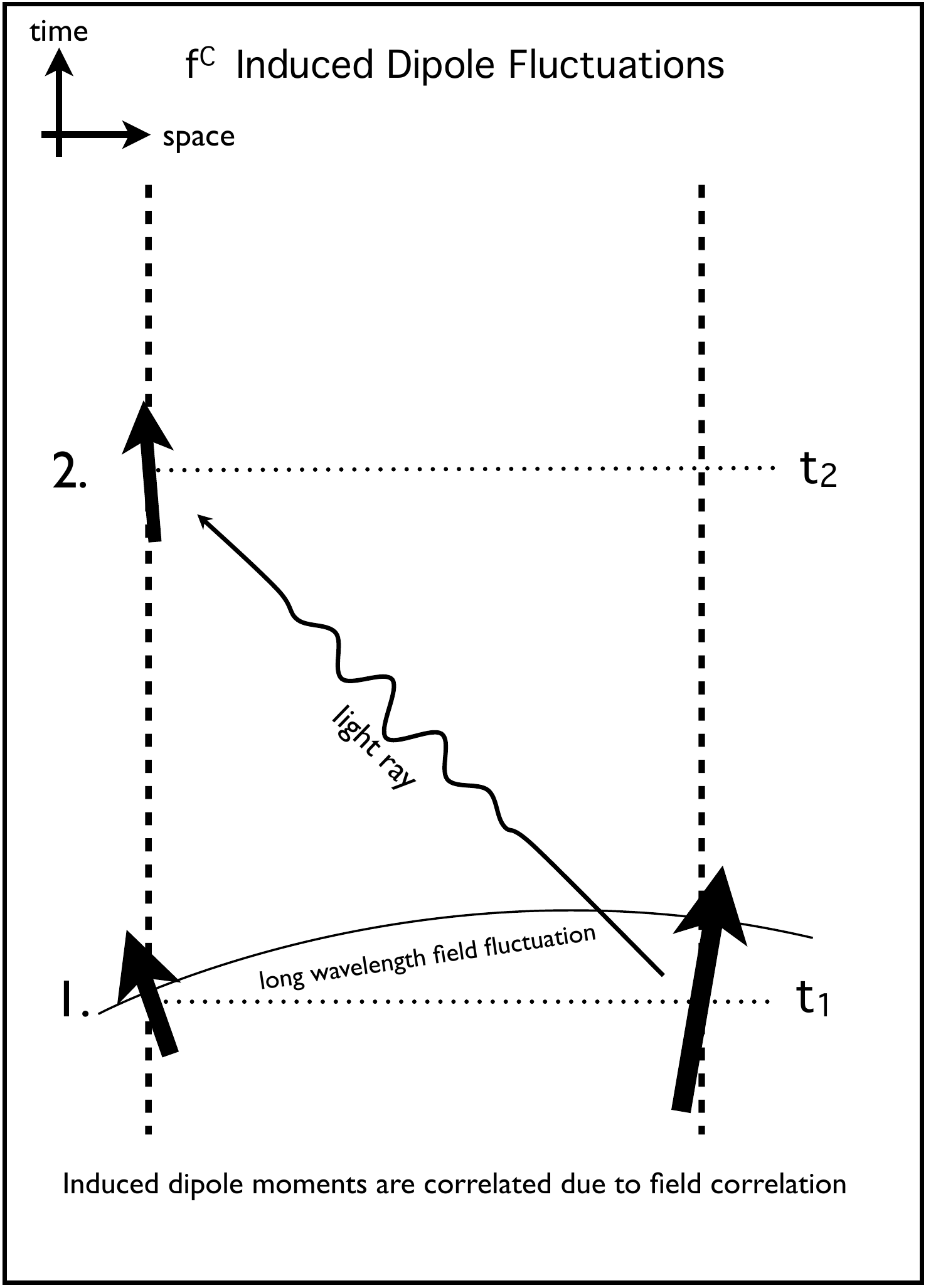} } }
\end{center}
\caption{
The illustrations depict the physical origin of the intrinsic fluctuation and induced dipole forces. \textit{On the left} intrinsic dipole fluctuations (represented by shaded oval); 1. radiate information about their motion, and 2. this radiation induces a correlated dipole moment in the second atom (solid black arrow denotes an induced dipole moment).
The induced motion at $t_2$ leads to radiation that travels back to the fluctuating atom. At $t_3$ the radiation produced at step 2 will produce a local electric field  near the fluctuating atom  which carries information about its own fluctuations in the past.
The illustration \textit{on the right} depicts the physical origin of the second component of the force arising from field fluctuations and their spatial correlation. Step 1 shows how a field fluctuation induces correlated dipole moments in both atoms. The induced motion of the dipole moments will lead to radiation emitted from both atoms containing information about their motion (only left moving radiation included). At $t_2$ the radiation generated by the induced motion produces a local electric field around each atom that is correlated with its motion.
} \label{fafb}
\end{figure}

The form of (\ref{fA}) and (\ref{fB}) can be explained by appealing
to the heuristic description of the force given in the introduction.
$f^A$ and $f^B$ arise from the $\it{intrinsic}$ $\it{fluctuations}$
in the dipole moments of the atoms. This can be seen by noting that they
contain the atom's Hadamard function i.e. the symmetric two point
function for the oscillator degree of freedom. The two retarded
electric field Green's functions account for the transfer of
information between the two atoms, and the retarded Green's function
for the atom characterizes its response to an external field see
Fig[\ref{fafb}].

The third component of the force, $f^C$ arises from $\it{induced}$ $\it{fluctuations}$
of the atom's dipole moments.
The retarded Green's functions for the two oscillators,  $g_{ret,a}$,
characterize their response to a given field fluctuation. The $k$th
component of the induced dipole moment of the $a$th atom can be
written as $d^k_{ind,a}= q_a \int d\lambda' g_{ret,a}(\lambda,
\lambda') E^k[z^\alpha_a(\lambda')]$ where
$E^k[z^\alpha_a(\lambda')]$ is the $k$th component of the electric
field at the position of the atom. The symmetric two-point function
of the induced dipole moment quantifies its fluctuations $ \left< \{
d^j_{ind,a}(t), d^k_{ind,b}(t') \} \right>= q_a q_b \int d\lambda
d\lambda'  g_{ret,a}(t, \lambda) g_{ret,b}(t', \lambda')
E^{jk}_H(z^\alpha_a(\lambda),z^\alpha_b(\lambda))$. The remaining
electric field propagator, $E^{ret}_{ij}$, carries information about
the motion of one atom to the other and accounts for the form of
$f^C$ see Fig[\ref{fafb}]

\begin{eqnarray}
\label{vdWC}
f^C_k(\tau)= \frac{1}{2} q_1^2 q_2^2 \int_{\lambda_i}^{\lambda_f} d\lambda \int_V d^4 x \int_V d^4 y \ g_{ret,1}(\tau, \lambda)[ \partial_k(x') E^{ij}_{ret}(x', x) G_{ret}(x,y)  E^{H}_{ij}(z^\alpha(\lambda), y) |_{x'=z^\alpha(\tau)} \nonumber \\ +
 E^{ij}_{ret}(z^\alpha(\tau), x) G_{ret}(x,y) \partial_k(x') E^{H}_{ij}(x', y) |_{x'=z^\alpha(\lambda)} ]
\end{eqnarray}
where $\partial_k(x)$ denotes differentiation with respect to
$x^{k}$. The previous form of the force is valid for any atomic
motion. However, a self consistent treatment would require that the
aforementioned  `self energy' terms be included in order to account
for the back-action of the field on the atom itself.

\subsection{Induced Dipole Force}

In this section we calculate the induced dipole force explicitly by
plugging in the retarded Green's function for the second oscillator,
$g_{ret,2}$, and choosing $\vec{z}_2$ to be the origin of
coordinates.
\begin{eqnarray}
\label{ }
f^C_k(\tau)= \frac{1}{2} q_1^2 q_2^2 \int_{\lambda_i}^{\lambda_f} d\lambda \int d t \int dt'  \ g_{ret,1}(\tau, \lambda)  g_{ret,2}(t,t') [ E^{ij}_{ret}(z^\alpha(\tau), t',\vec{0})  \partial_k(x') E^{H}_{ij}(x', t, \vec{0}) |_{x'=z^\alpha(\lambda)}  \nonumber \\ + E^{ij}_{H}(z^\alpha(\lambda), t',\vec{0})  \partial_k(x') E^{ret}_{ij}(x', t, \vec{0}) |_{x'=z^\alpha(\tau)}      ]
\end{eqnarray}
The derivatives operating on the various Green's functions can be
simplified by employing the equation of motion for the field, the
resultant form valid for general atom motion follows.

\begin{eqnarray}
\label{IDF}
f^C_k= \frac{q_1^2q_2^2}{4\pi}   \int d\lambda \int dt  \int d t' g_{ret,1}(\tau,\lambda) g_{ret,2}(t,t')   \theta(\tau,t')   \bigg\{
\delta^{'''}[\sigma(z^\alpha(\tau);t', \vec{0})] \bigg[
 \frac{1}{2} D_H^{''}[\sigma(z^\alpha(\lambda);t, \vec{0})]  \nonumber \\    \times
  \sigma(z^\alpha(\tau);t', \vec{0})_k
  \bigg(
      \sigma(z^\alpha(\lambda);t, \vec{0})_j \ \sigma(z^\alpha(\lambda);t, \vec{0})^j  \ \sigma(z^\alpha(\tau);t', \vec{0})_i \ \sigma(z^\alpha(\tau);t', \vec{0})^i   +    [ \sigma(z^\alpha(\lambda);t, \vec{0})_i \  \sigma(z^\alpha(\tau);t', \vec{0}) )^i ]^2   \bigg)
\nonumber \\ +
\delta^{''}[\sigma(z^\alpha(\tau);t', \vec{0})] \bigg[
 \frac{1}{2} D_H^{'''}[\sigma(z^\alpha(\lambda);t, \vec{0})]  \nonumber \\    \times
  \sigma(z^\alpha(\lambda);t, \vec{0})_k
  \bigg(
      \sigma(z^\alpha(\lambda);t, \vec{0})_j \ \sigma(z^\alpha(\lambda);t, \vec{0})^j  \ \sigma(z^\alpha(\tau);t', \vec{0})_i \ \sigma(z^\alpha(\tau);t', \vec{0})^i   +    [ \sigma(z^\alpha(\lambda);t, \vec{0})_i \  \sigma(z^\alpha(\tau);t', \vec{0}) )^i ]^2 \bigg)
 \nonumber \\
 + D_H^{''}[\sigma(z^\alpha(\lambda);t, \vec{0})]
  \bigg(  6  \sigma(z^\alpha(\tau);t', \vec{0})_i  \ \sigma(z^\alpha(\tau);t', \vec{0})^i
  \sigma(z^\alpha(\lambda);t, \vec{0})_k    +  2  \sigma(z^\alpha(\lambda);t, \vec{0})_i \ \sigma(z^\alpha(\tau);t', \vec{0}) )^i   \  \sigma(z^\alpha(\tau);t', \vec{0})_k \bigg)
     \bigg]
  \nonumber \\
     + \delta^{'}[\sigma(z^\alpha(\tau);t', \vec{0})] \bigg[
 4 \  D_H^{'''}[\sigma(z^\alpha(\lambda);t, \vec{0})]
  \sigma(z^\alpha(\tau);t', \vec{0})_i \ \sigma(z^\alpha(\tau);t', \vec{0})^i \
  \sigma(z^\alpha(\lambda);t, \vec{0})_k    \nonumber \\
 + 20 \ D_H^{''}[\sigma(z^\alpha(\lambda);t, \vec{0})]  \sigma(z^\alpha(\lambda);t, \vec{0})_k  \bigg]
   \bigg\} \ \ \
\end{eqnarray}
Here,  primes on functions denote derivatives with respect to
$\sigma$, and $\sigma_k=\partial_k \sigma$  denotes differentiation
of $\sigma$ with respect to $x^k$.

We can separate (\ref{IDF}) into 4 terms with differing number of $\sigma$-derivatives and specify a static trajectory for the distinguished atom to bring the derivatives outside of the integral i.e. $d/d\sigma=z^{-1}d/dz$. To distinguish which Green's function a given $\sigma$-derivative acts on we attach a dummy subscript to $z$ that should not be confused with an atom label. Once all derivatives are taken $z_1$ and $z_2$ are set to $z$, the separation between the two atoms.
The evaluation of the $t$-integral can be done by substituting $ \delta(\sigma(x,x'))=\frac{\delta(t'-t+|\vec{x}-\vec{x'}|)}{|\vec{x}-\vec{x'}|}$.

\begin{eqnarray}
\label{ }
f^{C1}_z(\tau)=-\frac{q_1^2q_2^2}{4\pi} z \  \frac{1}{2} \left[
  \left( \frac{1}{z_2}\frac{d}{dz_2} \right)^3   \left( \frac{1}{z_1}\frac{d}{dz_1} \right)^2
  +\left( \frac{1}{z_2}\frac{d}{dz_2} \right)^2   \left( \frac{1}{z_1}\frac{d}{dz_1} \right)^3 \right] \nonumber \\
     \frac{1}{z_1}  \int d\lambda \int dt \  g_{ret,1}(\tau,\lambda) g_{ret,2}(t,\tau-z_1)    D_H [\sigma(z_2^\alpha(\lambda),t, \vec{0})]   \bigg[ 2 z^4  \bigg]    \bigg|_{z_1=z_2=z}
\end{eqnarray}

\begin{eqnarray}
\label{ }
f^{C2}_z (\tau)=-\frac{q_1^2q_2^2}{4\pi} z
  \left( \frac{1}{z_2}\frac{d}{dz_2} \right)^2   \left( \frac{1}{z_1}\frac{d}{dz_1} \right)^2
     \frac{1}{z_1}  \int d\lambda \int dt \   g_{ret,1}(\tau,\lambda) g_{ret,2}(t,\tau-z_1)   D_H [\sigma(z_2^\alpha(\lambda),t, \vec{0})]      \bigg[  8 z^2 \bigg]   \bigg|_{z_1=z_2=z}
\end{eqnarray}

\begin{eqnarray}
\label{ }
f^{C3}_z(\tau)=-\frac{q_1^2q_2^2}{4\pi} z
  \left( \frac{1}{z_2}\frac{d}{dz_2} \right)^3   \left( \frac{1}{z_1}\frac{d}{dz_1} \right)
     \frac{1}{z_1}  \int d\lambda \int dt \  g_{ret,1}(\tau,\lambda) g_{ret,2}(t, \tau-z_1)    D_H [\sigma(z_2^\alpha(\lambda),t, \vec{0})]     \bigg[  4 z^2  \bigg]  \bigg|_{z_1=z_2=z}
\end{eqnarray}

\begin{eqnarray}
\label{ }
f^{C4}_z(\tau)=-\frac{q_1^2q_2^2}{4\pi} z
  \left( \frac{1}{z_2}\frac{d}{dz_2} \right)^2   \left( \frac{1}{z_1}\frac{d}{dz_1} \right)
     \frac{1}{z_1}  \int d\lambda \int dt \ g_{ret,1}(\tau,\lambda) g_{ret,2}(t, \tau-z_1  )     D_H [\sigma(z_2^\alpha(\lambda),t, \vec{0})]      \bigg[  20   \bigg]  \bigg|_{z_1=z_2=z}
\end{eqnarray}

We can express the Green's function  for the field through a mode sum
and subsequently evaluate the $\lambda$ and $t$ integrals in the long
time limit. The exact field-influenced dynamics of the oscillators will be
dissipative, however this dissipative effect does not  appear at this
order in perturbation theory, but can be modeled phenomenologically by inclusion of an infinitesimal
dissipation in the oscillator equation of motion i.e.
$g_{ret,a}(t-t') \rightarrow g_{ret,a}(t-t')e^{-\epsilon (t-t')}$.

At finite temperature the Hadamard function for the  field can be
obtained through periodicity in imaginary time, or by taking the
trace of symmetrized field operators with respect to a thermal
density matrix. Expressing the result as a mode sum we find.

\begin{eqnarray}
\label{ }
f^{C1}_z=-\frac{2}{\pi} z \  \frac{1}{2}
 \left[
  \left( \frac{1}{z_2}\frac{d}{dz_2} \right)^3   \left( \frac{1}{z_1}\frac{d}{dz_1} \right)^2
  +\left( \frac{1}{z_2}\frac{d}{dz_2} \right)^2   \left( \frac{1}{z_1}\frac{d}{dz_1} \right)^3 \right]    \nonumber \\
     \frac{1}{z_1 z_2}  \int_0^{\infty} d\omega \  \alpha_1(\omega) \alpha_2(\omega)   \coth (\beta \omega/2) \sin \omega z_2     \cos  \omega z_1   \bigg[  2 z^4  \bigg] \bigg|_{z_1=z_2=z}
\end{eqnarray}

\begin{eqnarray}
\label{ }
f^{C2}_z=-\frac{2}{\pi} z
  \left( \frac{1}{z_2}\frac{d}{dz_2} \right)^2   \left( \frac{1}{z_1}\frac{d}{dz_1} \right)^2
     \frac{1}{z_1 z_2}  \int_0^{\infty} d\omega \  \alpha_1(\omega) \alpha_2(\omega)   \coth (\beta \omega/2) \sin \omega z_2     \cos  \omega z_1   \bigg[  8 z^2  \bigg] \bigg|_{z_1=z_2=z}
\end{eqnarray}

\begin{eqnarray}
\label{ }
f^{C3}_z=-\frac{2}{\pi} z
  \left( \frac{1}{z_2}\frac{d}{dz_2} \right)^3   \left( \frac{1}{z_1}\frac{d}{dz_1} \right)
     \frac{1}{z_1 z_2}  \int_0^{\infty} d\omega \  \alpha_1(\omega) \alpha_2(\omega)   \coth (\beta \omega/2) \sin \omega z_2     \cos  \omega z_1   \bigg[  4z^2  \bigg] \bigg|_{z_1=z_2=z}
\end{eqnarray}

\begin{eqnarray}
\label{ }
f^{C4}_z=-\frac{2}{\pi} z
  \left( \frac{1}{z_2}\frac{d}{dz_2} \right)^2   \left( \frac{1}{z_1}\frac{d}{dz_1} \right)
     \frac{1}{z_1 z_2}  \int_0^{\infty} d\omega \  \alpha_1(\omega) \alpha_2(\omega)   \coth (\beta \omega/2) \sin \omega z_2     \cos  \omega z_1   \bigg[  20  \bigg] \bigg|_{z_1=z_2=z}
\end{eqnarray}
As the retarded Green's functions for the atoms characterizes the response of their dipole moments to an external field they play the role of the dynamic polarizability, $\alpha$.  The result above is expressed in terms of the frequency dependent form, $\alpha_a(\omega)= q_a^2(4\pi \mu_a(\Omega_a^2-\omega^2))^{-1}$ ,  which can be derived from the classical equations of motion (the aforementioned infinitesimal dissipative term kills an imaginary part in the infinite time limit). $\beta$ is the field's inverse temperature. All the derivatives can be taken and the force can be expressed as an integral over frequency.

\begin{eqnarray}
\label{ }
f^{C}_z =-\frac{2}{\pi z^7}
   \int_{0}^{\infty} d\omega \  \alpha_1(\omega) \alpha_2(\omega)   \coth (\beta \omega/2)
   \bigg[ \omega z ( 18-8z^2 \omega^2 +z^4 \omega^4) \cos 2\omega z  \nonumber \\ + (-9 +16 \omega^2 z^2 -3 \omega^4 z^4) \sin 2 \omega z \bigg]
   \end{eqnarray}
This expression agrees with what can be found in the literature for
the CP force in a finite temperature field \cite{Passante}. In the
far-field at zero temperature we recover the well known form \cite{CPP95}

\begin{equation}
\label{ }
f^C_z\approx - \frac{161}{4\pi} \alpha_1(0) \alpha_2(0) \frac{1}{z^8}
\end{equation}
where a UV regulator must be employed to render the frequency integrals finite.

If however we take the dissipation to be zero (as it truly is in our perturbative approach) the force is altered because the polarizability acquires an imaginary part $\alpha(\omega) \rightarrow (q^2/4\pi \mu \Omega)[\Omega/(\Omega^2-\omega^2)^{-1}+ i (\pi/2) \delta(\omega-\Omega)- i (\pi/2) \delta(\omega+\Omega)]$. The imaginary term plays an important role when the quantum nature of the dipole moment of the oscillator is accounted for. When such a term is neglected, the contribution to the atom-atom force from $f^A$ and $f^B$ dominates in the far field as $1/z^3$ rather than $1/z^8$ at $T=0$. We denote this contribution to the force by $\delta f^C_z$.

\begin{eqnarray}
\label{ }
\delta f^{C}_z =-\frac{q_1^2 q_2^2}{16 \pi^2 \mu_1 \mu_2 \ \Omega_1  \Omega_2 z^7}
   \int_{0}^{\infty} d\omega \   \coth (\beta \omega/2)
   \bigg[ \frac{\Omega_1 \delta(\omega-\Omega_2)}{\Omega_1^2-\Omega_2^2} + ( \Omega_1 \leftrightarrow \Omega_2) \bigg] \nonumber
   \\  \times \bigg[ -9-2 z^2\omega^2 -z^4 \omega^4 +(9-16 z^2\omega^2 +3 z^4 \omega^4)\cos (2\omega z)   + z\omega (18-8 z^2\omega^2 + z^4 \omega^4)\sin (2\omega z) \bigg]
\end{eqnarray}
As $z \rightarrow \infty$ this term possesses the asymptotic scaling,

\begin{eqnarray}
\label{ }
\delta f^{C}_z \approx \frac{q_1^2 q_2^2}{16 \pi^2 \mu_1 \mu_2 \Omega_1 \Omega_2 z^3}
   \int_{0}^{\infty} d\omega \   \omega^4  \coth (\beta \omega/2)
   \bigg[ \frac{\Omega_1 \delta(\omega-\Omega_2)}{\Omega_1^2-\Omega_2^2} +  (\Omega_1 \leftrightarrow \Omega_2) \bigg]
   \end{eqnarray}
but at $T=0$ for $z \rightarrow 0$ $\delta f^C$ is subleading to the dominant $1/z^7$ near field scaling from the London term.

\subsection{Intrinsic Fluctuation Force}

The treatment by London of the atom-atom force can be reproduced by
computing the interaction energy of two atoms interacting via the
Coulomb potential. The force follows from the negative gradient of
the perturbed energy eigenvalues.  We obtain an analogous expression
for the London force in our formulation but with an additional
contribution from retardation effects as we treat the field relativistically.                 

The contributions to the force from $f^A_z$ and $f^B_z$ can be computed in the same way as the contribution $f^C_z$, and so we omit the details of that calculation here, and only state the result in the long-time zero temperature limit.

\begin{equation}
\label{FA}
f^{A}_z=-\frac{q_1^2 q_2^2}{16 \pi^2 \mu_1 \mu_2 \Omega_1 \Omega_2} \frac{\Omega_1}{\Omega_1^2-\Omega_2^2}
[ 9+2 \Omega_2^2 z^2 +\Omega^4_2 z^4]
\frac{1}{z^7}
\end{equation}
$f^B_z$ can be obtained from $f^A_z$ by exchanging $\Omega_1$ and
$\Omega_2$. These terms are responsible for the near-field behavior
and agree with those derived by London when retardation corrections
to the field Green's function are neglected \cite{Lon}.

\begin{equation}
\label{ }
f^A_z+f^B_z \approx f^{Lon}_z= -\frac{9 q_1^2 q_2^2}{ 16 \pi^2 \mu_1 \mu_2 \Omega_1 \Omega_2} \frac{1}{\Omega_1+\Omega_2} \frac{1}{z^7}
\end{equation}

The thermal version of the  previous result does not make sense for a single oscillator where temperature is an ill-defined quantity, but does in the case of a gas of atoms. If the gas is sufficiently dilute the force between two collections of trapped atoms can be approximated using the density distribution of the gas and $f_z$ \cite{APSS08}.
The finite temperature form follows where $\beta_a$ is the inverse temperature of the $a$th oscillator (or trapped gas).

\begin{equation}
\label{}
f^{A}_z=-\frac{q_1^2 q_2^2}{16 \pi^2 \mu_1 \mu_2 \Omega_1 \Omega_2} \frac{\Omega_1}{\Omega_1^2-\Omega_2^2}
[ 9+2 \Omega_2^2 z^2 +\Omega^4_2 z^4] \coth( \beta_2 \Omega_2/2)
\frac{1}{z^7}
\end{equation}

In the far field the leading order behavior reduces to the following form.
\begin{equation}
\label{FA}
f^{A}_z \approx -\frac{q_1^2 q_2^2}{16 \pi^2 \mu_1 \mu_2} \frac{\Omega^3_2  }{\Omega_1^2-\Omega_2^2}
\coth( \beta_2 \Omega_2/2)
\frac{1}{z^3}
\end{equation}
Note that when the field and the atoms are in thermal equilibrium
this new asymptotic scaling cancels with an equal and opposite
contribution contained in $\delta f^C_z$ and the standard far field
scaling $1/z^8$ is restored. When the atoms and field are out of
thermal equilibrium this cancelation no longer occurs and the
dominant contribution to the force scales  like $1/z^3$, where the
zero temperature contribution cancels as indicated below:                              

\begin{equation}
\label{NS}
f_z \approx   -\frac{q_1^2 q_2^2}{8 \pi^2 \mu_1 \mu_2 } \frac{\Omega^3_2  }{\Omega_1^2-\Omega_2^2}
\bigg[ \frac{1}{e^{\beta_2 \Omega_2}-1}-\frac{1}{e^{\beta \Omega_2}-1}\bigg] \frac{1}{z^3} +(\Omega_1 \leftrightarrow \Omega_2)
\end{equation}

\section{Entanglement Force}

The previous derivation of the atom-atom force assumes that the initial state of the two oscillators is uncorrelated. If however, the two atoms are initially entangled then a new contribution to atom-atom force arises.
To our knowledge this force has not been reported in the literature.

We begin by computing the oscillator-reduced IF
for two initially entangled atoms.

\begin{eqnarray}
\label{orIF1E}
\mathcal{F}_Z[\vec{z}^-_1,\vec{z}^+_1]  =  \prod_a
\left[ \int d\vec{Q}_a d\vec{Q}_{in,a}   d\vec{Q}'_{in,a}
\int_{\vec{Q}_{in,a}}^{\vec{Q}_a}  \mathcal{D}\vec{Q}_a
\int_{\vec{Q}'_{in,a}}^{\vec{Q}_a}
  \mathcal{D}\vec{Q}'_a
e^{i(S_Q[\vec{Q}_a]-S_Q[\vec{Q}'_a])}
\right]
 \nonumber \\ \times
 \rho_Q(\vec{Q}_{in,1},\vec{Q}'_{in,1} ,\vec{Q}_{in,2},\vec{Q}'_{in,2} ;t_{in})  \mathcal{F}[J^{\mu-},J^{\nu+}]
\end{eqnarray}
Writing (\ref{orIF1E}) in a more
suggestive form

\begin{eqnarray}
\label{orIFE}
\mathcal{F}_Z[\vec{z}^{+}_1,\vec{z}^{-}_1]=  \mathcal{F}\left[\vec{z}^+_a, \vec{z}^-_a ; -i\frac{\delta}{\delta {\vec{j}_a}^+},-i\frac{\delta}{\delta {\vec{j}_a}^-}  \right]
F [\vec{j}_1^+,\vec{j}_1^-,\vec{j}_2^+,\vec{j}_2^-] \bigg|_{{j_a}^\pm=0}
\end{eqnarray}
defines the IF for two entangled harmonic oscillators, $F [\vec{j}_1^+,\vec{j}_1^-,\vec{j}_2^+,\vec{j}_2^-] $.
To bring $\mathcal{F}[J^{\mu-},J^{\nu+}]$ out of the path integrals in (\ref{orIF1E}) we replace the oscillator coordinates with functional derivatives as before.
\begin{eqnarray}
\label{foE}
F [\vec{j}_1^+,\vec{j}_1^-,\vec{j}_2^+,\vec{j}_2^-] = \prod_a \int d{\vec{Q}_a } \int  d{\vec{Q}_{in,a}} d{\vec{Q}_{in,a}}  \ \rho_{Q}(\vec{Q}_{in,1},\vec{Q}'_{in,1} ,\vec{Q}_{in,2},\vec{Q}'_{in,2} ; t_{in} )  \nonumber  \\ \int_{\vec{Q}_{in,a}}^{\vec{Q}_a} \mathcal{D}\vec{Q}_a\int_{\vec{Q}_{in,a}}^{\vec{Q}_a} \mathcal{D}\vec{Q}'_a  e^{iS_Q[\vec{Q}_a]-iS_Q[\vec{Q}'_a]+i\int d\lambda [\vec{j}_a^+\cdot\vec{Q}_a^-+\vec{j}_a^-\cdot\vec{Q}_a^+]}
\end{eqnarray}
For the initially entangled squeezed Gaussian state

\begin{eqnarray}
\label{ }
\rho_Q(\vec{Q}_{in,1},\vec{Q}'_{in,1} ,\vec{Q}_{in,2},\vec{Q}'_{in,2} ;t_{in})= \left( \frac{\beta}{\pi \alpha} \right)^6 \exp \bigg\{ -\frac{1}{4} \bigg[ \beta^2
\left(    (\vec{Q}_{in,1}+\vec{Q}_{in,2})^2 + (\vec{Q}'_{in,1}+\vec{Q}'_{in,2})^2  \right)
\nonumber \\
+ \frac{1}{\alpha^2}
\left(    (\vec{Q}_{in,1}-\vec{Q}_{in,2})^2 + (\vec{Q}'_{in,1}-\vec{Q}'_{in,2})^2  \right) \bigg] \bigg\}
\end{eqnarray}
(\ref{foE}) can be evaluated exactly. For this case, like oscillator coordinate components are entangled together with equal magnitude in each direction i.e. the parameters $\alpha$ and $\beta$ are common to each component.

The influence functional for two entangled oscillators follows

\begin{eqnarray}
\label{ }
F [\vec{j}_1^+,\vec{j}_1^-,\vec{j}_2^+,\vec{j}_2^-]  = \tilde{\mathcal{N} } \exp \bigg\{ \frac{1}{8} \bigg(
i B_{\Delta} g_\Delta+ i B_{\Sigma} g_\Sigma-\alpha^2 B_\Delta^2-\frac{1}{\beta^2} B_\Sigma^2-\frac{1}{\alpha^2} g_\Delta^2-\beta^2 g_\Sigma^2 +i \varphi \bigg)
\bigg\}
\end{eqnarray}
where $\tilde{\mathcal{N}}$ is a normalization constant  and the definitions of $B_a$, $g_a$ and $\varphi$ are defined below.

\begin{equation}
\label{ }
B_a= \int dt \ J^-_a(t) [\cot \Omega T \sin \Omega(t-t_i)+\csc \Omega T \sin \Omega(t_f-t) ]
\end{equation}
\begin{equation}
\label{ }
g_a=\frac{1}{\mu \Omega} \int dt \  J^-_a(t) \sin \Omega(t-t_i)
\end{equation}
\begin{equation}
\label{ }
\varphi=\sum_a \bigg[ \frac{1}{\sin \Omega T} \int dt \  g_a J'_a(t) \sin \Omega(t_f-t) -\frac{1}{2} \mu \Omega g_a^2 \cot \Omega T +\frac{i}{2} \int dt dt' [J_a(t) J_a(t') g^a_F(t,t')+J'_a(t) J'_a(t') g^a_D(t,t') ] \bigg]
\end{equation}
The subscript $\Sigma$ denotes $C_\Sigma=C_1+C_2$, and the subscript $\Delta$ denotes $C_\Delta=C_1-C_2$.

 The entanglement force comes from the leading order contribution to $\mathcal{F}_Z$ that depends upon the spatial separation between the atoms. Previously we needed to consider the square of $S_{cross}$. However, when the two atoms are entangled there exists nonvanishing cross correlation between their coordinates such that $\left< S_{cross} \right>_o \neq 0$.

So in distinction to the previous section we have

\begin{eqnarray}
\label{ }
\mathcal{F}_Z  =  e^{iS_{IF}[z^\pm]}  \approx  1+(\text{self energy terms})+i \left< S_{cross}[\vec{z}_a^\pm,\vec{Q}_a^\pm]\right>_o +\mathcal{O}({z^-}^2)
\end{eqnarray}
where the force can be derived from

\begin{equation}
\label{EF}
f^{ent}_k(\tau) \approx -\frac{\delta }{\delta z_1^{k-}(\tau)}\left< S_{cross}[\vec{z}^\pm_a,\vec{Q}^\pm_a] \right>_o \bigg|_{z^{k-}=0}.
\end{equation}

Expanding $S_{cross}$ for small $z^{k-}_1$ we arrive at

\begin{eqnarray}
\label{ }
S_{cross} \approx S_o+ q_1 q_2 \int d\lambda d\lambda' z_1^{k-}(\lambda) [
 \partial_k \kappa_i^\mu  \kappa^{\nu'}_j  D^{ret}_{\mu \nu}( z^{\alpha+}_1(\lambda),  z^{\alpha}_2(\lambda')) Q^{i+}_1(\lambda)   Q^{j+}_2(\lambda') \nonumber \\
 +
 \frac{1}{4} \partial_{k'} \kappa_i^{\mu'}  \kappa^{\nu}_j  D^{ret}_{\mu \nu}( z^{\alpha}_2(\lambda),  z^{\alpha +}_1(\lambda')) Q^{i-}_1(\lambda)   Q^{j-}_2(\lambda')
 +
 \frac{i}{2} \partial_{k} \kappa_i^{\mu}  \kappa^{\nu'}_j  D^{H}_{\mu \nu}( z^{\alpha +}_1(\lambda),  z^{\alpha}_2(\lambda')) Q^{i+}_1(\lambda)   Q^{j-}_2(\lambda') ]  +  \mathcal{O}({z^-}^2)
\end{eqnarray}
where a prime in the index of a derivative operator means
differentiation with respect to the second argument. Only one term
survives after we take the expectation value,  that which contains
the cross correlator $\left<  Q^{i+}_1(\lambda)   Q^{j+}_2(\lambda')
\right>_o$ which equals

\begin{equation}
\label{ } \left<  Q^{i+}_1(t)   Q^{j+}_2(t')  \right>_o=  \delta^{ij}
\ g_{ent}(t,t') = \frac{1}{4}\bigg(\frac{1}{\alpha^2}-\beta^2\bigg)\
\delta^{ij} \ \bigg[ \frac{1}{(\mu\Omega)^2} \sin \Omega(t-t_i) \sin
\Omega(t'-t_i)- \frac{\alpha^2}{4 \beta^2} \csc^2 \Omega T S(t) S(t')
\bigg]
\end{equation}
where $T=t_f-t_i$ and $S(t)= \sin \Omega(t-t_f)-\sin
\Omega(t-t_i+T)$. After taking the expectation value and then using
(\ref{EF}) we obtain the entanglement force.

\begin{equation}
\label{ }
f^{ent}_k(\tau)= - q_1 q_2  \int d\lambda \ g_{ent}(\tau,\lambda)
 \partial_k \kappa_i^\mu  \kappa^{i \nu'}  D^{ret}_{\mu \nu}( z^{\alpha}_1(\tau),  z^{\alpha}_2(\lambda))
\end{equation}
All derivatives can be taken on the field's retarded Green's function and simplified using the equation of motion. We then specify the trajectory to be static to arrive at

\begin{equation}
\label{ }
\partial_k \kappa^\mu_i \kappa^{i \nu} D^{ret}_{\mu \nu}(\sigma)= \delta_{k z} \left[ z^3 \left( \frac{d}{d \sigma} \right)^3 + 5 z \left( \frac{d}{d \sigma} \right)^2 \right] D_{ret}(\sigma).
\end{equation}
With a static trajectory specified the $\sigma$ derivatives can be
expressed  in terms of $z$ derivatives and can be factored out of the
integral. The retarded Green's function can be expressed as a delta
function $D_{ret}(\sigma(z_1^\alpha(\tau),
z_2^\alpha(\lambda)))=\delta(\tau-\lambda-z)/(4\pi z)$.  We work
in a coordinate system centered at the $a=2$ atom, where various
forces act upon (also the origin of the xy-plane), with $z$ axis
along the ray connecting the two atoms at distance $z$ apart
(pointing from atom 2 to atom 1). Thus the distinguished atom ($a=1$)
is located at ($0, 0, z$).                                              
 This leads to the explicit expression for the entanglement force.

\begin{equation}
\label{ } f^{ent}_z(\tau)= -\frac{q_1 q_2}{4 \pi} \left[ z^3 \left(
\frac{1}{z} \frac{d}{d z} \right)^3 + 5 z \left( \frac{1}{z}
\frac{d}{d z} \right)^2 \right] \frac{1}{z} g_{ent}(\tau,\tau-z)
\end{equation}

In the infinite time limit $\tau \rightarrow t_f \rightarrow \infty$ $g_{ent}(\tau,\tau-z) \sim \frac{1}{8}( \beta^2-1/\alpha^2)(\alpha^2/\beta^2- 1/(\mu \Omega)^2) \cos \Omega z$. The force vanishes in the far field but has a well-defined near field limit i.e. $\Omega z \rightarrow 0$.

\begin{equation}
\label{EFf}
f^{ent}_z \sim - \frac{q_1 q_2}{32 \pi} \left( \beta^2-\frac{1}{\alpha^2} \right)\left(\frac{\alpha^2}{\beta^2}- \frac{1}{(\mu \Omega)^2} \right) \frac{\Omega^2}{z^2}
\end{equation}

This effect is not only due to entanglement between the two atoms, but it is also due to retardation.
For the case considered above where the degree of entanglement between like components of the atom's dipole moments has the same magnitude the interaction energy as described through the Coulomb potential vanishes. Thus, only through the inclusion of relativistic effects does any force manifest.

However, the previous discussion can easily be generalized to the case where the magnitude of the parameters $\alpha$ and $\beta$ is not common to all directions. As such the initial state for the oscillators takes the generalized form

\begin{eqnarray}
\label{ }
\rho_Q(\vec{Q}_{in,1},\vec{Q}'_{in,1} ,\vec{Q}_{in,2},\vec{Q}'_{in,2} ;t_{in})= \prod_j  \left( \frac{\beta_j}{\pi \alpha_j}  \right)^2 \exp \bigg\{ -\frac{1}{4} \bigg[ \beta_j^2
\left(    (Q^j_{in,1}+Q^j_{in,2})^2 + (Q^{j'}_{in,1}+ Q^{j'}_{in,2})^2  \right)
\nonumber \\
+ \frac{1}{\alpha_j^2}
\left(    (Q^j_{in,1}-Q^j_{in,2})^2 + (Q^{j'}_{in,1} -Q^{j'}_{in,2})^2  \right) \bigg] \bigg\}.
\end{eqnarray}
The development follows closely that given for the previous case and so we only state the result in the near-field long-time limit,

\begin{equation}
\label{EFcoul}
f^{ent}_z\approx -\frac{3}{4\pi} q_1q_2 \bigg[ \Delta_x +\Delta_y
-2\Delta_z  \bigg] \frac{1}{z^4}
\end{equation}
where we have used the shorthand $\Delta_j= \frac{1}{8}( \beta_j^2-1/\alpha_j^2)(\alpha_j^2/\beta_j^2- 1/(\mu \Omega)^2) $.
Note that if the parameters $\alpha$ and $\beta$ are equal for all directions (\ref{EFcoul}) vanishes. The sign of the force can also be changed by the appropriate choice of the squeeze parameters $\alpha$ and $\beta$.

\section{Possibility of Detection}

\subsection{Atom and Field Out of Thermal Equilibrium}

In this section we compute the relative magnitude for the atom-atom force when the field and atoms are not in thermal equilibrium to the force at zero temperature. We focus our attention on the case where the atom's are in their ground state and the field is in a thermal state of inverse temperature $\beta$.
Measuring this new asymptotic scaling requires a balance between temperature and the first optical resonance of the atomic species used. For the case when $\Omega \beta>>1$ (\ref{NS}) is exponentially suppressed, this would rule out the use of heavier atoms like Rb near room temperature, the only hope is to work in the regime where $\beta \Omega \gtrsim 1$, not only to prevent suppression by the Planck factor but also  to prevent the excitation of the atom so that measurement can be done before thermalization.

The relative magnitude of (\ref{NS}) to $f_C$ in the far field shows when this new scaling will dominate. For realistic experiments, atom-atom distance of the order of $\mu m$, the high temperature limit is beyond access for the temperatures and atomic species we are considering, so we replace $\delta f^C$ with its zero temperature form (also the appropriate factors of $c$ have been restored to ensure that (\ref{Ratio}) is dimensionless).

\begin{equation}
\label{Ratio}
\frac{\delta f^C}{f^C} =- \frac{8\pi}{161 c^5} \frac{\Omega_1^2 \Omega_2^5}{\Omega_1^2-\Omega_2^2} z^5 \frac{1}{e^{\beta \Omega_2}-1} +( \Omega_1 \leftrightarrow \Omega_2)
\end{equation}
If the atomic species are the same the previous expression reduces to

 \begin{equation}
\label{ }
\frac{\delta f^C}{f^C} =\frac{24\pi}{322 } \left( \frac{\Omega z }{c} \right)^5 \frac{1}{e^{\beta \Omega}-1}.
\end{equation}
 Tuning $\Omega$ to hydrogen's first optical resonance ($\Omega \approx 10 \text{eV}$,  $\Omega \approx 2.4 \times
 10^{15} \text{Hz}$, or $\Omega \approx 116, 000 \text{K} $ ) we find
 \begin{equation}
\label{ }
\frac{\delta f^C}{f^C} \approx \frac{24\pi}{322 } \left( \frac{8 z}{\mu m} \right)^5 \frac{1}{e^{\beta \Omega}-1}.
\end{equation}

If the atomic species are different we find different behavior. Particularly, when one of the atom's first optical resonance is very large such that $\beta \Omega>>1$ (like Rb near room temperature) the Planck factor for that atom will be strongly suppressed so its contribution to the force can be ignored. In such a case (\ref{Ratio}) takes the form

\begin{equation}
\label{}
\frac{\delta f^C}{f^C} =- \frac{8\pi}{161} \left( \frac{\Omega z }{c} \right)^5 \frac{1}{e^{\beta \Omega}-1} =- \frac{8\pi}{161}     \left( \frac{8 z}{\mu m} \right)^5
\frac{1}{e^{\beta \Omega}-1}
\end{equation}
where we have a different sign and a slightly different coefficient.

For atoms $ \delta f^C/f^C$ only becomes significantly greater than 1 for large distances and very high temperatures and so is unlikely observable. However,
these effects may play a role in the laboratory
for molecules with sub $eV$ excitation energies. We leave a study of those effects for a later work.


\subsection{Entanglement Force}

Now that we have an expression for the entanglement force at short distances we check for regimes in which (\ref{EF}) will dominate. To do this we take the ratio of the entanglement force to the near-field van der Waals force. After restoring all physical constants to yield the correct dimensions and allowing both atoms to be the same species we find.

\begin{equation}
\label{ }
\frac{f^{ent}}{f^{Lon}}=\frac{4 \epsilon_o}{9 c^2} \frac{\mu \Omega^4}{q^2} \left( \tilde{\beta}^2-\frac{1}{\tilde{\alpha}^2} \right) \left( \frac{\tilde{\alpha}^2}{\tilde{\beta}^2} -1 \right) z^5
\end{equation}
Above $\tilde{\beta}=\beta/\mu\Omega$ and $\tilde{\alpha}=\mu\Omega \alpha$.

\begin{figure}
\begin{center}
\includegraphics[width=4in]{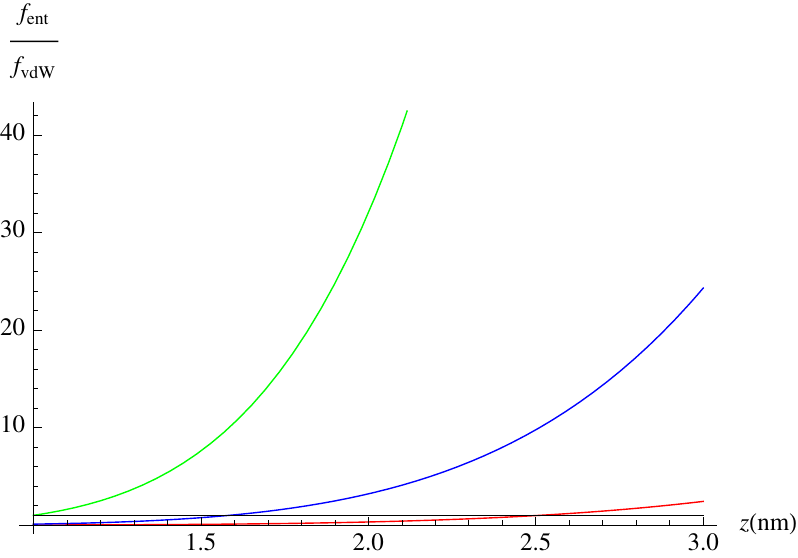}
\caption{Relative magnitude of the entanglement force to the London force. Intersection of the various graphs with the black line yields the value of the interatomic distance where the the two forces are equal in magnitude. The various colors represent different values for the prefactor in (\ref{OOM}) i.e. $ P=8.9 \times  \left( \tilde{\beta}^2-\frac{1}{\tilde{\alpha}^2} \right) \left( \frac{\tilde{\alpha}^2}{\tilde{\beta}^2} -1 \right)$. For red $P=0.01$ for blue $P=0.1$ and for green $P=1$. The plot shows that for interatomic distances satisfying the near field condition $\Omega z<<1$ that the entanglement force dominates for order $1$ prefactor for distances of a few nanometers. }
\label{POD}
\end{center}
\end{figure}

By tuning the frequency to the first optical resonance of Hydrogen, taking the reduced mass to be the electron mass and $q$ to be the electronic charge we find.

\begin{equation}
\label{OOM}
\frac{f^{ent}}{f^{Lon}}  \approx 8.9 \times  \left( \tilde{\beta}^2-\frac{1}{\tilde{\alpha}^2} \right) \left( \frac{\tilde{\alpha}^2}{\tilde{\beta}^2} -1 \right)   \left( \frac{z}{  \text{nm} } \right)^5
\end{equation}
The near field condition requires that the distance between the atoms be much smaller than the wavelength associated with their first optical resonance. For hydrogen this wavelength is $\lambda= c /\Omega =122 \text{nm} $. So, for the case where the prefactor  of (\ref{OOM}) is order unity and the interatomic distances are in the range of a few nanometers we find the entanglement force dominates over the standard London form see Fig[\ref{POD}].

\subsection{Conclusion}

In this paper we have laid down the theoretical groundwork for the
study of interatomic forces under fully nonequilibrium conditions. As
a first step we have employed the influence functional formalism to
derive a fully dynamical description of the atom-atom force for
general atomic motion and initial states. We have found that a
careful treatment of the infinite time limit shows the existence of a
novel far field scaling when the atoms and field are not in thermal
equilibrium. The dominance of this term in the laboratory would
require a careful balance between temperature and the first optical
resonance of the atomic species.
%
%
For entangled atoms a novel near-field scaling is
obtained that dominates the standard London force in certain regimes.
These new forces could play an important role in quantum computing schemes involving entangled atoms.

\end{document}